\documentclass[journal,draftclsnofoot,onecolumn,12pt]{IEEEtran} 

\IEEEoverridecommandlockouts

\usepackage{times}
\usepackage{amsmath,amssymb}
\usepackage{graphicx}
\usepackage{algorithmic}
\usepackage{algorithm}
\usepackage{color}
\usepackage{cite}
\usepackage{graphicx,epstopdf}
\usepackage[algo2e]{algorithm2e}
\usepackage{bm}
\usepackage[keeplastbox]{flushend}
\usepackage{multirow,hhline}

\usepackage[justification=centering]{caption}

\newcommand{\CASE}[1]{\STATE \textbf{case} #1\textbf{:} \begin{ALC@g}}
\newcommand{\ENDCASE}{\end{ALC@g}}

\newcommand{\DEFAULT}{\STATE \textbf{default:} \begin{ALC@g}}
\newcommand{\ENDDEFAULT}{\end{ALC@g}}
\newcommand{\DEFAULTLINE}[1]{\STATE \textbf{default:} }

\begin{document}

\bibliographystyle{IEEEtran}

\title{Scheduling for VoLTE: Resource Allocation Optimization and Low-Complexity Algorithms}
\author{
Maryam Mohseni$^*$, S. Alireza Banani, Andrew W. Eckford, and Raviraj S. Adve%
\thanks{$^*$Corresponding author, e-mail: mmohseni81@yahoo.com}%
\thanks{M. Mohseni and S. A. Banani were, and R. S. Adve is, with the Dept. of ECE,
University of Toronto.}%
\thanks{A. W. Eckford is with the Dept. of EECS, York University, 4700 Keele Street,
Toronto, Ontario, Canada M3J 1P3.}%
\thanks{Financial support for this work was provided by TELUS and the Natural Sciences and Engineering Research Council.}
}

\maketitle 

\vspace*{-10ex}
\begin{abstract}
We consider scheduling and resource allocation in long-term evolution (LTE) networks across voice over LTE (VoLTE) and best-effort data users. The difference between these two is that VoLTE users get scheduling priority to receive their required quality of service. As we show, strict priority causes data services to suffer. We propose new scheduling and resource allocation algorithms to maximize the sum- or proportional fair (PF) throughout amongst data users while meeting VoLTE demands. Essentially, we use VoLTE as an example application with both a guaranteed bit-rate and strict application-specific requirements. We first formulate and solve the frame-level optimization problem for throughput maximization; however, this leads to an integer problem coupled across the LTE transmission time intervals (TTIs). We then propose a TTI-level problem to decouple scheduling across TTIs. Finally, we propose a heuristic, with extremely low complexity. The formulations illustrate the detail required to realize resource allocation in an implemented standard. Numerical results show that the performance of the TTI-level scheme is very close to that of the frame-level upper bound. Similarly, the heuristic scheme works well compared to TTI-level optimization and a baseline scheduling algorithm. Finally, we show that our PF optimization retains the high fairness index characterizing PF-scheduling.
\end{abstract}


\section{Introduction}
\label{Sec:intro}
Long-term evolution, or LTE, was proposed by the 3rd generation partnership project (3GPP) to address the continually increasing demands for high data rates and ubiquitous connectivity. Importantly, LTE allows for a diverse set of mobile applications such as high definition television (HD TV), online gaming, video meetings, etc. A crucial aspect of LTE is that, in order to achieve high data rates with myriad user applications, the available radio resources are efficiently allocated among several users~\cite{4}. This allocation is achieved using orthogonal frequency division multiple access (OFDMA) and is based on user requirements, current system load and system configuration. 

LTE networks only support packet services, and to be supported, circuit-switched services must migrate to be packet-switched. One important example of this is voice. It is widely expected that, over the next few years, mobile voice services will increasingly migrate toward Voice over LTE (VoLTE)~\cite{2}. Importantly, for our purposes, VoLTE places strict quality of service (QoS) constraints each VoLTE user must satisfy within the packet-switched framework.

Broadly, LTE defines two QoS categories: guaranteed bit rate (GBR) services, of which VoLTE is one example, and non-GBR services. GBR services have priority in the allocation of available time-frequency resources; non-GBR services are best-effort, receiving any remaining resources. Table~\ref{table:QoS} lists the QoS Class of Identifier (QCI), and QoS constraints, for different LTE services~\cite{15}.

\begin{table*}[t!]
  \centering
   \caption{QoS class of identifier (QCI)~\cite{15}}\label{table:QoS}
  \begin{tabular}{|c|c|c|c|c|c|}
    \hline
    {\bfseries QCI} & {\bfseries Bearer Type} & {\bfseries Priority} & 
    	{\bfseries Packet Delay} & {\bfseries Packet Loss} & {\bfseries Example} \\
    \hhline{------}
    1 &  \multirow{4}{*}{GBR} & 2 & 100 ms & 10$^{-2}$ & VoIP call \\
    \hhline{-~----}
    2 & & 4 & 150 ms & \multirow{2}{*}{10$^{-3}$} & Video call \\
    \hhline{-~--~-}
    3 & & 3 & 50 ms & & Online Gaming (Real Time) \\
    \hhline{-~----}
    4 & & 5 & 300 ms & \multirow{3}{*}{10$^{-6}$} & Video streaming \\
    \hhline{----~-}
    5 & \multirow{5}{*}{Non-GBR} & 1 & 100 ms & & IMS Signaling \\
    \hhline{-~--~-}
    6 & & 6 & 300 ms & & Video, TCP based services e.g. email, chat, ftp, etc. \\
    \hhline{-~----}
    7 & & 7 & 100 ms & 10$^{-3}$ & Voice, Video, Interactive gaming \\
    \hhline{-~----}
    8 & & 8 & \multirow{2}{*}{300 ms} & \multirow{2}{*}{10$^{-6}$} & \multirow{2}{*}{Video, TCP based services e.g. email, chat, ftp, etc.} \\
    \hhline{-~-~~~}
    9 & & 9 & & & \\
    \hhline{------}
  \end{tabular}
  \vspace{4pt}
 \end{table*}

The LTE scheduler is tasked with satisfying GBR users~\cite{13}. For VoLTE, the constraints state that an LTE user must receive a fixed number of coded bits in every 20ms frame, no more, no less (300 coded bits for our example choice of encoder\footnote{Throughout this paper, we use representative numbers for various allowed LTE and VoLTE parameters. Other VoLTE and LTE implementation choices would change these numbers, but not the problem formulation or solution methodology.}). Furthermore, these bits must be delivered in exactly one of the 20 transmission time intervals (TTIs) in each frame. This relatively low rate, coupled with the strict priority given GBR users and the granularity allowed in LTE frequency allocations, could significantly reduce the overall spectral efficiency of the LTE network. Consequently, we need effective resource allocation to enhance the throughput performance of data users while satisfying the GBR users as much as possible. For our purposes, an effective algorithm is one that accounts for the specifics of the application at hand. 

\subsection{Contributions}

There are several works proposing resource allocation for OFDMA networks with various metrics~\cite{PFCoMP,NetworkCoded,MultipleChoice,Ref1,Ref2}. For example,~\cite{Ref1} presents a mathematical framework for studying the capacity of OFDM multi-cell networks while multiple services with different QoS constraints exist (aiming to maximize the number of admissible users). Also~\cite{Ref2} considers minimizing the total transmit power with power control while mutual interference among multiple cells exist. This paper is closely tied to the LTE system rather than a generic OFDM/OFDMA system; the central point of this paper is that, when optimizing throughput in real-world systems, the details of the application(s) play a vital role. An optimization problem with a generic framed OFDM system would either have to include the details of the framing, TTIs, bits per frame, the CQI indicator system etc. represented as a variables (resulting a large number of variables, but still only applicable to a single system) or be removed from the real-world system. As GBR users become more popular and more applications migrate to using LTE, we believe our optimization framework will become increasingly important. 

Our main contributions are:
\begin{itemize}
\item We formulate the frame-level rate-maximization problem while accounting for the details of the VoLTE transmission scheme. Since the VoLTE users require a fixed data rate, the maximization is effectively over the non-GBR users. The VoLTE constraints couples scheduling across TTIs and leads to integer constraints coupled across TTIs. This makes any solution extremely computationally expensive. Moreover, in some scenarios, e.g., low bandwidth or high number of VoLTE users, it is impossible to serve all VoLTE users and the frame-level problem is infeasible and the problem does not provide any allocation solution.
\item To reduce the computation load and, in infeasible scenarios, provide service to as many VoLTE users as possible, we propose a two-phase TTI-level optimization problem: the first sub-problem selects the VoLTE users to be scheduled in a TTI to satisfy as many VoLTE users as possible. The second sub-problem then allocates LTE physical resource blocks (PRBs) to the chosen users in order to maximize the throughput for the best-effort non-GBR users. While still an integer programming problem, the solution space is far smaller. 
\item We develop and solve the proportional fair (PF) version of this two-phase problem implemented as maximizing the weighted sum rate of the data users.
\item In order to further reduce the computational complexity, we propose a simple heuristic scheme, with low computational complexity, to solve the two-phase sum-throughput and sum PF throughput optimization problems. As our numerical results show, the heuristic performs extremely close to the TTI-level scheme (the upper bound for TTI-level scheduling). Moreover, it outperforms the baseline scheduling algorithm, especially in low bandwidth and dense networks. We also show that the solution for PF rate maximization retains the fairness that characterizes PF scheduling.
\end{itemize}

\subsection{Background and Related Works}
\label{sec:relwork}

Scheduling is one of the most important components in LTE systems: the scheduler, implemented at the eNodeB (eNB), distributes the limited available time/frequency resources among the active users in order to satisfy their QoS needs~\cite{1,16}. The 3GPP specifications do not specify scheduling policies, making the scheduling algorithm vendor-specific~\cite{17}. There are several works that have proposed and compared various scheduling algorithms; however, in general, these do not use an optimization framework to maximize a chosen metric (such as our choice of sum rate or PF rate) while meeting constraints set by GBR users. In this paper we focus on VoLTE users; the principles developed here could be extended to other GBR flows.

\begin{figure}[htp]
\centering
\includegraphics[totalheight=26em]{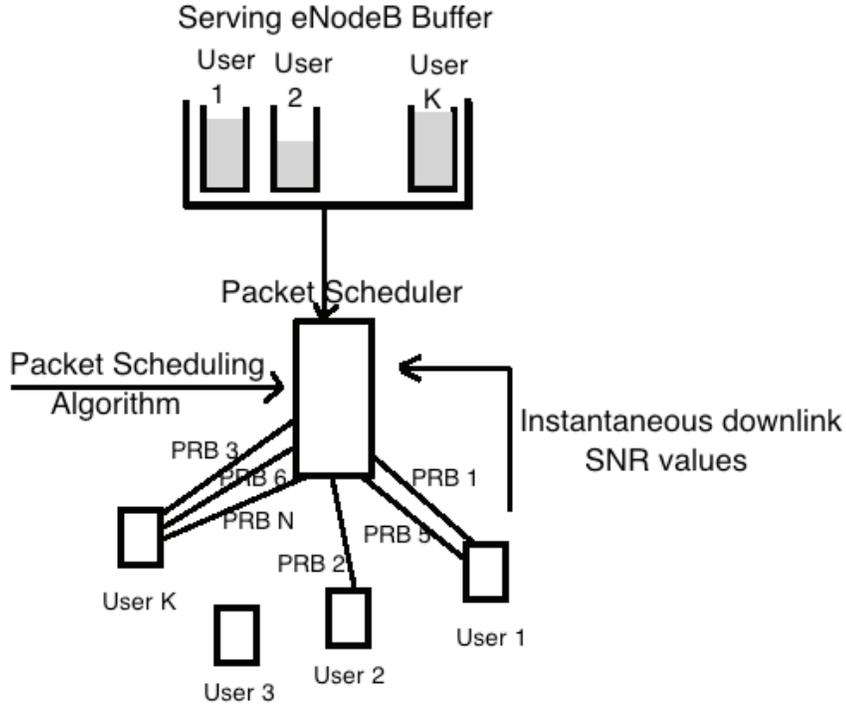}
\caption{Generalized model of scheduling in the downlink of an LTE system. Recreated from~\cite{4}}\label{fig:LTEsch}
\end{figure}

Fig.~\ref{fig:LTEsch} illustrates a generalized model of packet scheduling for LTE systems in the downlink. Users feed back their received signal-to-interference-plus-noise ratio (SINR) to the eNB. The scheduler uses these values, combined with the users' needs and knowledge of previous allocations, to determine which user should be scheduled on which PRB. The SINR are related to channel quality indicators (CQIs) which determine the modulation and coding scheme (MCS) used. In this paper, we assume that the CQI is available at the granularity of a PRB. An LTE scheduler comprises a time-domain scheduler (selecting the users to be served according to QoS requirements) and a frequency-domain scheduler (allocating the resources to the users selected by the time-domain scheduler). 

The three basic scheduling algorithms are round robin, max sum rate, and proportional fair~\cite{3}. The round robin scheduler assigns the PRBs in equal TTIs and in ordered manner. It allocates resources equally, regardless of the users' CQI. The max-sum rate scheduler maximizes throughput by assigning resources to users that have the highest SINR, but ignores fairness among users. The proportional fair scheduler provides a balance between fairness and the overall system throughput. Compared to the sum rate objective, achieving rate allocations with PF provides a better tradeoff between users' satisfaction (i.e., fairness) and total system throughput~\cite{PF}. 

These three basic scheduling algorithms are modelled and compared in~\cite{7} and~\cite{8} in terms of both throughput and fairness index. In~\cite{Compare}, the authors compare six scheduling schemes including round robin, max rate, Kwan max throughput, MaxMin, PF, and resource fair (RF) via simulations. In~\cite{3} a new scheduling mechanism is proposed to increase the throughput compared to PF and round robin schedulers, while losing only 20\% of fairness among all users. While not named, this algorithm uses the users' CQIs and past allocations to determine the allocations in the next TTI. Also, in~\cite{SCH_NFair} two scheduling algorithms are proposed which are more complex than max rate scheduling, but provide fair scheduling among users. 

For multi-tier networks a scheduling algorithm based on PF scheduling and frequency reuse is developed in~\cite{MTier}. The work in~\cite{PFCoMP} develops resource allocation for Coordinated Multi-Point transmission in LTE-Advanced to ensure a good tradeoff between cell-average and cell-edge spectral efficiency. Most of these works consider non-GBR users only (while we propose novel scheduling algorithms which consider two different services non-GBR and GBR services).

Recently, there has been considerable interest in real-time multimedia services such as video and Voice-over-IP (VoIP). In~\cite{1} the performance of three promising scheduling algorithms, FLS, EXP rule and LOG rule are compared. In~\cite{9}, the performance of different scheduling algorithms such as PF and EXP/PF are evaluated and compared for different flows such as best-effort and video. A new scheduling algorithm based on a utility function, called UBS, is proposed in~\cite{UBS} proposing separate PF scheduling for different types of services. These works consider each application separately while our work considers both best-effort and VoLTE services jointly. Also, in~\cite{GTReview} game theory based resource allocation algorithms for LTE networks are presented while~\cite{GT} proposes scheduling based on class service using cooperative game theory. The available resources are fairly distributed among classes in proportion, which results in higher fairness levels amongst classes. The users with tightest delay requirements are prioritized.

Specifically for VoLTE,~\cite{2} proposes a resource allocation algorithm based on the prediction of channel quality change for VoLTE users. Also, in~\cite{Ref4} a semi-persistent packet scheduling algorithm is proposed for VoLTE (GBR) users over heterogenous networks of fourth generation LTE and third-generation universal mobile telecommunications service (3G UMTS). A new channel- and QoS-aware scheduling algorithm (WE-MQS) is proposed in~\cite{Ref3} for downlink scheduling of VoLTE users which considers user perception. To reduce the waste of resources when scheduling a low-rate VoLTE user on a strong channel, a scheduling algorithm is proposed in~\cite{MuxVoice} in which voice packets for different users are multiplexed within one LTE packet in the downlink. Compared to these works we consider both GBR and non-GBR users in LTE networks. Also, the authors of~\cite{RBPreserver} proposed a Resource Block Preserver scheduling algorithm for both real time and non-real time flows. It includes two layers, the upper layer is  designed to satisfy the QoS of real time flows based on sub-frame aggregation, while the lower layer allocates the resource blocks to users depending on flow types. In~\cite{MixTr} and~\cite{DLQA} new scheduling algorithms are developed which can manage mixed traffic including both real time and non-real time traffic. 

In LTE networks, providing QoS for multimedia services is important and the QoS parameters in terms of throughput, packet loss, delay and fairness are analyzed and evaluated for different scheduling schemes; however the quality of experience, related to end user experience, is not as well studied. In~\cite{QoE1} the performance of three scheduling schemes is evaluated in terms of QoE for VoIP services. Also, in~\cite{QoE2} a new QoE-driven LTE downlink scheduling algorithm is proposed for VoIP applications and compared to existing mechanisms.
  
Clearly, there are several proposed algorithms (and evaluations) for scheduling in LTE systems without GBR QoS guarantees, and a few with GBR QoS guarantees. The motivation behind our work comes from two different aspects: first, most available scheduling algorithms for LTE do not consider GBR services (e.g., VoLTE), whereas we consider LTE scheduling for two classes of services (non-GBR and GBR). Second, instead of giving strict high priority to GBR users, which causes data users to suffer and reduces overall cell throughput, we approach this problem from an optimization point of view, to use available resources efficiently, while satisfying QoS of GBR users. In what follows we develop and solve several optimization problems that account for the details of the VoLTE constraints.

\subsection{Organization of Paper}
This paper is organized as follows: Section~\ref{sec:system} presents the system model and an example to illustrate our motivation. Section~\ref{sec:optFr} develops a frame-level optimization problem to maximize cell throughput, our first contribution. Section~\ref{sec:optTS} presents two TTI-level optimization problems to maximize total throughput and, optionally, achieving proportional fairness. Section~\ref{sec:heuristic} then presents the different low-complexity heuristic algorithms for different objectives. Section~\ref{sec:results} presents results of simulations that illustrate the performance of proposed scheduling schemes. Finally, Section~\ref{sec:conc} concludes the paper with some discussion of our approaches.

\section{System Model}
\label{sec:system}

\begin{figure}[htp]
\centering
\includegraphics[totalheight=14em]{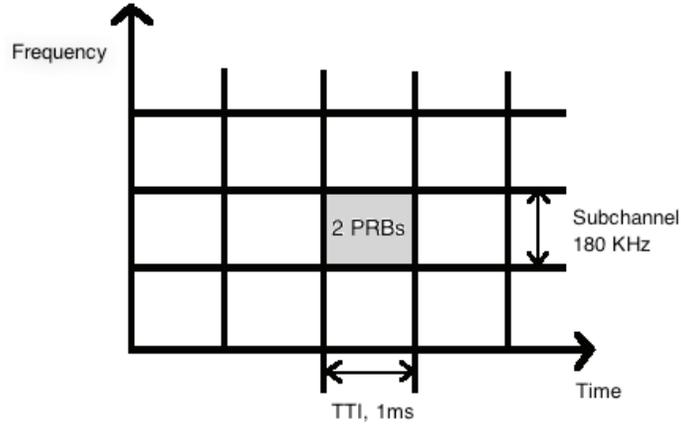}
\caption{Radio resources in time and frequency in LTE.}\label{fig:PRB}
\end{figure}

We begin with a brief review of LTE, as applied to our problem. Radio resources in the downlink of an LTE system, called PRBs, specify a time-frequency allocation. A single PRB comprises 12 consecutive sub-carriers (for a total bandwidth of 180 kHz) and either 6 (extended cyclic prefix) or 7 (normal cyclic prefix; we assume a normal cyclic prefix) OFDM symbols in one time slot of 0.5 ms. As shown in Fig.~\ref{fig:PRB}, two consecutive time slots together form one TTI~\cite{6}. Resource allocation occurs at the granularity of 12 subcarriers and one TTI, i.e., 14 OFDM symbols. We note that in the literature (and in our system model below), a `PRB' often refers to a block of 12 subcarriers with the time dimension of one TTI separated out.

We consider a single LTE eNodeB serving $U$ VoLTE users and $K$ data users, i.e., overall we have $U+K$ users in the cell. The eNB has $T$ TTIs, indexed by $t=1,2,\ldots,T$ (frame of $T$ TTIs for downlink transmissions excluding uplink transmissions), and $N$ PRBs, indexed by $n=1,2,\ldots, N$, available for all downlink transmissions. Each user feeds back its CQI for each of the $N$ PRBs. If a user is allocated a specific PRB, the eNB chooses the appropriate MCS according to the CQI reported by the user. Table~\ref{fig:MCS} lists the CQI and the SINR switching threshold for the corresponding MCS that satisfies a 10\% Block Error Rate (BLER)~\cite{18}. 

\begin{table}[h]
\centering
\caption{CQI index, SINR switching threshold and corresponding MCS for 10\% BLER~\cite{18}}\label{fig:MCS}
\begin{tabular}{|c|c|c|c|c|}
\hline
{\bfseries CQI} & {\bfseries Modulation} & {\bfseries Code Rate }& \multirow{2}{*}{{\bfseries $\beta$}} & {\bfseries SINR threshold} \\
{\bfseries Index} & {\bfseries Order} & {\bfseries $\times$ 1024} & & {\bfseries (dB)}\\
\hline
0 & \multicolumn{4}{c|}{No transmission} \\
\hline
1 & QPSK & 78 & 1.00 & -9.478 \\
\hline
2 & QPSK & 120 & 1.40 & -6.658 \\
\hline
3 & QPSK & 193 & 1.40 & -4.098 \\
\hline
4 & QPSK & 308 & 1.48 & -1.798 \\
\hline
5 & QPSK & 449 & 1.50 & 0.399 \\
\hline
6 & QPSK & 602 & 1.62 & 2.424 \\
\hline
7 & 16QAM & 378 & 3.10 & 4.489 \\
\hline
8 & 16QAM & 490 & 4.32 & 6.367 \\
\hline
9 & 16QAM & 616 & 5.37 & 8.456 \\
\hline
10 & 64QAM & 466 & 7.71 & 10.266 \\
\hline
11 & 64QAM & 567 & 15.5 & 12.218 \\
\hline
12 & 64QAM & 666 & 19.6 & 14.122 \\
\hline
13 & 64QAM & 772 & 24.7 & 15.849 \\
\hline
14 & 64QAM & 873 & 27.6 & 17.786 \\
\hline
15 & 64QAM & 948 & 28.0 & 19.809 \\
\hline
\end{tabular}
\vspace{4pt}
\end{table}

We assume VoLTE is configured to use the Adaptive Multi-Rate Wideband (AMR-WB) 12.65 coder, and Robust Header Compression (RoHC) is enabled. The AMR-WB 12.65 coder generates 253 bits of coded speech every 20ms (a net rate of 12.65 kbps)~\cite{AMR-WB}. In addition, the Radio Link and Multiple Access Control layers add overhead, so the air interface must transport roughly 300 bits every VoLTE packet, in a time frame of 20ms or 20 TTIs (other VoLTE and LTE implementation choices would change these numbers, but not our problem formulation or solution methodology). Importantly, all these bits must be transmitted in \emph{exactly one} of the 20 TTIs in the 20ms frame (for downlink scheduling), i.e, these bits \emph{cannot be spread across multiple TTIs}. 

While one PRB (12 subcarriers, 14 symbols) includes 168 Resource Elements (REs), after accounting for overhead reference signals, approximately 120 REs/PRB per ms are available~\cite{PRB1,PRB2}. Each RE carries 2, 4 or 6 coded bits based on the CQI. For example, if CQI index is 15, the coding rate is 0.926 and each RE can carry 6 bits. As a result, each RE holds 6*0.926=5.55 data bits on average, and each PRB can then carry 120*5.55=666 data bits. Since LTE cannot allocate less than one PRB to each user, in this example \emph{one PRB will be allocated to one VoLTE packet}, wasting a significant amount of data space. With the same procedure, a single PRB can carry 177 bits and 18 bits for CQI 7 and CQI 1 respectively. These values are equivalent to 2 and 16 PRBs for one VoLTE packet in cases of CQI 7 and CQI 1 respectively. 

\section{Frame-level Optimization}
\label{sec:optFr}

The example above illustrates the need for a new scheduling algorithm. An LTE eNB must, usually, serve both VoLTE and data users. Because VoLTE users have a critical constraint on delay, they (traditionally) receive higher scheduling priority. Given the granularity of the allowed scheduling, as described above, a simplistic strict priority-based scheduler (our baseline scheme) would be inefficient. In a strict priority based scheduler, no data user can be scheduled even though there are not enough PRBs for a VoLTE user to be scheduled while satisfying QoS, thereby wasting PRBs. In turn, if the system must support a large number of VoLTE users, data users will suffer considerably. It is therefore important to develop a resource allocation scheme to not only ensure the QoS of VoLTE users, but also reduce their influence on other data services. Our main contribution in this section is the formulation of the frame-level optimization problem.

Our objective is to maximize total cell throughput including both VoLTE and data users. However, as in frame-level optimization problem we assume that all VoLTE users are scheduled and, since VoLTE users receive a fixed rate (every VoLTE user receives 300 bits within the 20ms frame), our objective function only considers the data users. Each user experiences different SINRs in different PRBs. So, based on the reported CQI, and the MCS table in Table~\ref{fig:MCS}, we can calculate the number of bits that each PRB can carry for each user, denoted by $B_{n,u}$. 

Recalling that $U$ is the number of VoLTE users and $K$ is the number of data users, the set of users $\{1,2,\ldots,U+K\}$ is arranged so that users $\{1,\ldots,U\}$ are VoLTE users, and users $\{U+1,\ldots,U+K\}$ are data users. We define a set of binary variables, $X_{n,u,t}$ : when $X_{n,u,t}=1$, PRB $n$ is assigned to user $u$ at TTI $t$, and $X_{n,u,t}=0$ otherwise. The objective function, the throughput of data users, is then given by
\begin{equation} 
R = \sum_{t=1}^{T} \sum_{k=1}^K \sum_{n=1}^N  X_{n,U+k,t} B_{n,U+k},\label{eq:rate}
\end{equation}
where, since the VoLTE constraint is over 20 TTIs, we set $T=20$.

We maximize this objective function subject to the three constraints. The first constraint is that each PRB can be assigned to only one user in any TTI. Therefore, we have
\begin{align}
\label{eq:f1} \sum_{u=1}^{U+K}  X_{n,u,t} = 1, \,\,\,\,\,\, 1 \le n \le N,  \,\, 1 \le t \le T.
\end{align}

The next constraint enforces the condition that each VoLTE user can be scheduled only once during each 20ms cycle. (We note that a general sum-rate maximization problem, independent of a standard, would not consider such a constraint.) To define this constraint, we introduce a set of auxiliary binary variables for the VoLTE users, $Y_{u,t}$. $Y_{u,t} = 1$ indicates that VoLTE user $u$ is scheduled in TTI $t$. Thus, we have
\begin{align}
\label{eq:f2} \sum_{t=1}^{T}  Y_{u,t} = 1,  \,\, \,\, \,\, \,\, 1 \le u \le U.
\end{align}

Finally, each VoLTE user must receive at least 300 bits (number of bits per VoLTE packet) when it is scheduled, but zero bits otherwise. This imposes the $U \times T$ conditions
\begin{align}
\label{eq:f3} \sum_{n=1}^N  X_{n,u,t} B_{n,u} \geq 300\times Y_{u,t}, \,\,\,\,\,\, 1 \le u \le U,  \,\, 1 \le t \le T.
\end{align}
This implies that if $Y_{u,t}=0$ for VoLTE user $u$, then we have the constraint $\sum_{n=1}^N  X_{n,u,t} B_{n,u} \geq 0.$ In this situation, based on the objective of maximizing the sum rate and constraints in~(\ref{eq:f1})-(\ref{eq:f2}), only the equality will hold (i.e., $X_{n,u,t}=0$ for all PRBs and VoLTE user $u$). This allows another VoLTE or data user to use the corresponding PRBs by allowing $X_{n,u',t}=1$ for some $u' \neq u$. 

On the other hand, if $Y_{u,t}=1$ for some VoLTE user $u$, then the following constraint holds
\begin{align}
\label{eq:f5} \sum_{n=1}^N  X_{n,u,t} B_{n,u} \geq 300,
\end{align}
which ensures that if a VoLTE user is scheduled, the user is allocated enough PRBs such that the entire VoLTE packet can be transmitted, i.e., the user's QoS constraint is satisfied.

Based on the above analysis, using \eqref{eq:rate}-\eqref{eq:f5}, we can formulate the optimization problem as 
\begin{eqnarray}
\label{eq:obj1}
  \mbox{P1:} &&  \max_{{\bf X},{\bf Y}} \sum_{t=1}^{T} \sum_{k=1}^K \sum_{n=1}^N  X_{n,U+k,t} B_{n,U+k} \\
 \label{eq:cons1} \mbox{s.t.}   &&   \sum_{u=1}^{U+K}  X_{n,u,t} = 1, \forall n,t\\
  \label{eq:cons2}  &&   \sum_{t=1}^{T}  Y_{u,t} = 1, \forall u \in \{1,\ldots,U\}\\
  \label{eq:cons3}  &&  \sum_{n=1}^N  X_{n,u,t} B_{n,u} \geq 300 \times Y_{u,t}, \forall t, u \in \{1,\ldots,U\} \\    
         && X_{n,u,t} \in \{0,1\}, \forall n,u,t\\
         && Y_{u,t} \in \{0,1\}, \forall t, u \in \{1, \label{eq:cons4}\ldots,U\}
  \end{eqnarray}

Problem P1 optimizes scheduling across both time and frequency simultaneously, covering the entire frame, i.e., the output indicates in each TTI which users should be served on which PRB. Since the optimization variables are binary, this problem is an Integer Linear Problem (ILP)~\cite{ILP1}. However, ILPs are known to be NP-hard and with exponential execution time, prohibitive even for reasonable problem sizes. One approach is to recognize that after relaxation, the problem is a linear program (LP) and enforce the binary constraints after solving the LP.

An additional complication in problem P1 is that it includes joint optimization over a frame of $T=20$ TTIs. This is because the VoLTE specifications state that each VoLTE user must receive 300 bits (including overhead) each 20ms. Furthermore, each such user must receive the entire block of 300 bits within one TTI. This coupling of the allocation variables across TTIs makes solving the optimization problem significantly more complex. 

The frame-level optimization problem assumes that the CQI remains constant for the entire 20 TTI frames. It also suffers from another, potentially fatal, drawback: it maximizes the data users' throughput while \emph{requiring that all VoLTE users are served}. In some scenarios, e.g., low bandwidth or high number of VoLTE users, the problem is infeasible and returns no solution. 

We address these issues in the next section; specifically, we decouple the scheduling in different TTIs and reduce solution complexity by developing a TTI-level problem in the sense of maximum total throughput while satisfying VoLTE constraints. In the case of an infeasible problem (frame-level), the TTI-level solution serves as many VoLTE users as possible and always converges.

\section{TTI-Level Optimization}
\label{sec:optTS}
For a scheme to be practical it must be executable in reasonable time and provide a reasonable sub-optimal solution when the original problem is infeasible. In this section we formulate the resource allocation problem for two different objectives: maximizing the overall throughput and the PF throughput among non-GBR users.
 
\subsection{Maximizing Total Throughput}

The previous frame-level objective was to maximize data users throughput while all VoLTE users are served. The constraint in~\eqref{eq:cons2} coupled the solution across TTIs. Here, we formulate an alternative, two-phase, TTI-level optimization problem while yet satisfying the VoLTE constraints. At any given TTI $t$, the first priority is to assign the available resources to VoLTE users in order to maximize total throughput of VoLTE users. As each VoLTE user has a predefined data rate, the first phase is equivalent to satisfying the maximum number of VoLTE users. Then, the second objective is to allocate resources to data users in a way that maximizes the sum throughput of these users. We now remove the subscript $t$ from the notation, since all variables are for a single TTI. However, the decisions made for the first TTI will impact on the next TTI (as explained later). We formulate an optimization problem for each of the two phases.

For the first optimization problem, we try to maximize VoLTE users' throughput or equivalently maximize the number of scheduled VoLTE users in the current TTI. So, the objective function for the first phase can be formulated as $\sum_{u=1}^U Y_{u}$ while $\{Y_{u}\}_{u=1}^U$ are the optimization variables. As before, the optimization must meet the constraint, similar to~(\ref{eq:cons1}), that each PRB can only be allocated to a single user in any TTI, i.e., $\sum_{u=1}^{U+K}  X_{n,u} = 1, n = 1, \dots, N$. We also, have the constraint similar to~\eqref{eq:cons3}
\begin{align}
\label{eq:f7} \sum_{n=1}^N  X_{n,u} B_{n,u} \geq 300 \times Y_{u}, u \in \{1,\ldots,U\},
\end{align}
though its interpretation is now slightly different. If $Y_{u}=0$, then we have the following constraint,
\begin{align}
\label{eq:f8} \sum_{n=1}^N  X_{n,u} B_{n,u} \geq 0.
\end{align}
In previous frame-level optimization problem because the objective was to maximize throughput of data users, if a VoLTE user is not scheduled in a current TTI, the PRBs would be allocated to other VoLTE users or data users. However, in this optimization problem the PRBs may be allocated to a VoLTE user even if it is not scheduled. This is because the optimization function does not depend on the PRB allocations. This is not a desirable allocation.

A similar problem arises if $Y_{u}=1$; we then have the following constraint 
\begin{align}
\label{eq:f9} \sum_{n=1}^N  X_{n,u} B_{n,u} \geq 300,
\end{align}
which, as before, enforces that if the VoLTE user is scheduled, it should transmit the entire VoLTE packet (at least 300 bits). However, again, in the previous frame-level optimum solution, when the number of PRBs were enough for one VoLTE packet, the remaining PRBs would be used to maximize the throughput of data users (objective function).  However, in this problem extra PRBs may be given to VoLTE users which is, again, not desirable.

In summary, based on the above analysis for~(\ref{eq:f7}), on solving the optimization problem, we can trust the outcomes in $\{Y_u\}$ only, but not $\{X_{n,u}\}$. In our two-phase approach, therefore, the first phase is to identify the optimum choice of VoLTE users to service in a specific TTI ($Y_u$). Putting together the objective and the constraints we have the following optimization problem
\allowdisplaybreaks
\begin{eqnarray}
\label{eq:obj2}
  \mbox{P2(a):} &&  \max_{{\bf X},{\bf Y}} \sum_{u=1}^U Y_{u} \\
 \label{eq:cons5} \mbox{s.t.}   &&   \sum_{u=1}^{U+K}  X_{n,u} = 1, \forall n\\
  \label{eq:cons6}  &&  \sum_{n=1}^N  X_{n,u} B_{n,u} \geq 300 \times Y_{u}, \forall u \in \{1,\ldots,U\}\\    
 \label{eq:cons7}
         && X_{n,u} \in \{0,1\}, \forall n,u \\
           && Y_u \in \{0,1\}, \forall u \in \{1,\ldots,U\}
    \end{eqnarray}

After solving optimization problem P2(a), we have identified which VoLTE users are to be scheduled in the current TTI (the non-zero $Y_u$), but the PRB allocation requires a second phase, as described next.

The second phase addresses the original problem of throughput maximization. In this optimization problem, we need to maximize the throughput of data users with the knowledge of which VoLTE users are scheduled in the current TTI ($Y_u$ is known from the first phase, problem P2(a)). The optimization variables are, therefore, $X_{n,u}$. Putting together the objective and the constraints we have 
\allowdisplaybreaks
\begin{eqnarray}
\label{eq:obj3}
  \mbox{P2(b):} &&  \max_{{\bf X}} \sum_{k=1}^K \sum_{n=1}^N  X_{n,U+k} B_{n,U+k} \\
 \label{eq:cons8} \mbox{s.t.}   &&   \sum_{u=1}^{U+K}  X_{n,u} = 1, \forall n\\
  \label{eq:cons9}  &&  \sum_{n=1}^N  X_{n,u} B_{n,u} \geq 300 \times Y_{u}, \forall u \in \{1,\ldots,U\}\\    
 \label{eq:cons10}
         && X_{n,u} \in \{0,1\}, \forall n,u
\end{eqnarray}

The two consecutive optimization problems, P2(a) and P2(b), are solved for each TTI in the 20-TTI frame. The solutions for both are feasible (both algorithms converge). However, it is important to note that neither problem includes the 'single use' VoLTE constraint of~\eqref{eq:cons2}. To implement this constraint, in every frame of $T=20$ TTIs, we begin with the pool of all VoLTE users. Then, in each subsequent TTI, the set of VoLTE users is updated by \emph{removing} the VoLTE users which were scheduled in a previous TTI. 

In practice, this means that the early TTIs in a frame are dominated by VoLTE users while later TTIs are dominated by data users. The number $U$ which, in the previous frame-level optimization, included all VoLTE users, in problem P2 reduces from one TTI to the next. In other words, in each TTI, $U$ indicates the  number of VoLTE users \emph{remaining to be served} in that frame. This number resets to the total number of VoLTE users at the end of each 20ms frame.

We define $C_{volte}$ and $C_{data}$ as the total amount of data transmitted to VoLTE users and data users till the current TTI. These parameters will be updated at the end of each TTI as follows
\begin{align}
\label{eq:f10}C_{volte}& = C_{volte}+253 \times \sum_{u=1}^U Y_{u},\\
\label{eq:f11} C_{data} & = C_{data}+\sum_{k=1}^K \sum_{n=1}^N  X_{n,U+k} B_{n,U+k}.
\end{align}

It is worth noting that P2(a) and P2(b) remain linear integer problems, though now not explicitly coupled across TTIs.  

\subsection{Achieving PF Throughput}

In the previous subsection, the objective of the scheduling was to maximize the total throughput. The proposed scheme can efficiently use available network resources, but may result in data users with good channel conditions starving other data users; in most service scenarios, this is not desirable. In order to achieve fairness two different metrics could be considered: achieving absolute fairness (maximizing minimum throughput) or achieving proportional fairness (PF). In this paper, we consider PF for the long-term average throughput of data users. PF has been shown to be equivalent to maximizing a logarithmic utility function of the \emph{long-term} transmission rate over all users~\cite{PF}; to convert this long-term average into a TTI based resource allocation, we use the fact that PF is equivalent to maximizing the weighted sum rates with the weights updated after each TTI depending on the average rate received by each user~\cite{Multi}. 

Consider a general network with $K$ users. Channel time is divided into discrete TTIs. The throughput of user $k$ at TTI $t$ is $C_{k,t}$. In each TTI, $\overline{A}_k(t+1)$ represents a weighed average rate for user $k$ up to TTI $t$, and is given by,
\begin{equation}
\label{eq:pf1}\overline{A}_k(t+1) = \gamma \overline{A}_k(t) + (1-\gamma) {C}_{k,t},
\end{equation}
where $0< \gamma <1$ is a parameter that balances the weights of the throughput in the past and the most recent TTI. When $\gamma$ is small, a higher weight is given to the current transmission rate, as shown in~(\ref{eq:pf1}), and the scheduling reacts to channel condition changes quickly. In this case, fairness can be achieved in a relatively large time scale. When $\gamma$ is large, the effect of current channel condition on the scheduling is reduced. Meanwhile, fairness is achieved over a smaller time period as $\gamma$ increases. Proportional fair scheduling can be achieved through iterative scheduling, where, in each TTI, we obtain scheduling decisions to maximize the objective $\sum_{k=1}^K C_{k,t}/\overline{A}_k(t)$ so the weight for each user $k$ would be $1/{\overline{A}_k(t)}$.

In our problem, the first priority must remain the satisfaction of the VoLTE users' requirements, i.e., the fairness is only amongst the data users. Giving the VoLTE users priority implies that the first optimization problem is the same as P2. We then allocate the remaining resources to data users in a way that maximizes PF throughput. In this second phase, the equivalent weighted sum rate problem can be formulated as
\allowdisplaybreaks
\begin{eqnarray}
\label{eq:obj3}
  \mbox{P2(c):} &&  \max_{{\bf X}} \sum_{k=1}^K \sum_{n=1}^N  \frac{X_{n,U+k} B_{n,U+k}} {\overline{A}_k(t)} \\
 \label{eq:cons8} \mbox{s.t.}   &&   \sum_{u=1}^{U+K}  X_{n,u} = 1, \forall n\\
  \label{eq:cons9}  &&  \sum_{n=1}^N  X_{n,u} B_{n,u} \geq 300 \times Y_{u}, \forall u \in \{1,\ldots,U\}\\    
 \label{eq:cons10}
         && X_{n,u} \in \{0,1\}, \forall n,u.
\end{eqnarray}
Note that, as in the previous section, the parameter $U$ indicates the VoLTE users scheduled in that specific TTI.

The basic idea is that at each TTI, the scheduling priority is given to the data users with the highest $\frac{C_{k,t}} {\overline{A}_k(t)}$. That is, data users that either can achieve high throughput in the current TTI (i.e., high $C_{k,t}$) or did not receive much in the past (i.e., low $\overline{A}_k(t)$) have a better chance to be scheduled in the current TTI. 

The problem in P2(c) is solved TTI by TTI. At each TTI $t$, $\overline{A}_k(t)$ can be calculated based on information before $t$. At the end of TTI $t-1$ or beginning of TTI $t$, $\overline{A}_k(t)$ is updated as
\begin{equation}
\label{eq:Amt}
  \overline{A}_k(t) = \left\{ \begin{array} {l}
    \gamma \overline{A}_k(t-1) + (1-\gamma) C_{k,t-1}, \\
      \ \ \ \ \ \ \ \ \ \ \ \ \ \mbox{if data user $k$ is scheduled in TTI $t-1$} \\
   \gamma \overline{A}_k(t-1), \ \  \mbox{otherwise.}
\end{array}
 \right.  
\end{equation}

As in the previous subsection, with the objective of maximizing overall throughput, after solving problems P2(a) and P2(c), the set of potential VoLTE users to be scheduled is updated. Also, $C_{volte}$ and $C_{data}$ are updated as in~(\ref{eq:f10}) and~(\ref{eq:f11}).

For both objectives (maximizing total throughput and achieving PF), in each TTI we first determine the VoLTE users to be scheduled (time-domain scheduler) and then data users to be served are selected and resource allocation to all selected users is done (time and frequency scheduling). Importantly, the fact that P2(a), P2(b), and P2(c) remain integer problems, motivates the search for a heuristic.

\section{Low-complexity Heuristic Scheduling Schemes}
\label{sec:heuristic}

In the previous two sections we developed frame-level and TTI-level resource allocation problems to maximize the throughput and the PF throughput. While the TTI-level two-phase approach combining problems P2(a) and P2(b) (or P2(a) and P2(c) for the PF case) eliminates the coupling across TTIs, these are still integer problems with the attendant computational complexity. It is, therefore, likely that these problems serve as performance benchmarks, but are not executable in real-time. 

In this section we propose two heuristic scheduling schemes with extremely low complexity; these algorithms are to be executed on a TTI by TTI basis. The TTI-level formulation in the previous section was, therefore, essential to proposing the heuristic. At each TTI, the process must decide which PRB should be assigned to which user so both user selection and PRB allocation (time and frequency-domain scheduling) would be done simultaneously. The first heuristic is proposed in order to maximize the overall throughput and the second achieves PF throughput among different data users.

\subsection{Maximizing Total Throughput}
We define two sets, $\mathcal{U}$ and $\mathcal{K}$: $\mathcal{U}$ denotes the set of all VoLTE users that have \emph{not been} scheduled until the current TTI (initially $\mathcal{U}$ includes all VoLTE users $\mathcal{U}=\{1,2,\ldots,U\}$). This set is updated from one TTI to the next using the results of the scheduling process. The set $\mathcal{K}=\{1,2,\ldots,K\}$ denotes the set of all data users.

\begin{algorithm}{}
\caption{Heuristic Scheme for maximizing overall throughput} 
\label{alg:1}
\begin{algorithmic}[1]
\STATE Initialize: $n=1$
\WHILE {$n \leq N$} 	
	\STATE Initialize: $D_{PRBs}=0$, $D_{PRB}=0$, $n_{PRBs}=0$
	\IF {$\mathcal{U} \ne \emptyset$}
		\STATE Find $u^*=\arg\max_{u\in\mathcal{U}} B_{n,u}$ 
		\STATE $D_{PRBs}=D_{PRBs} + B_{n,u^*}$ 
		\STATE $n_{PRBs}=n_{PRBs} + 1$
		\WHILE {$D_{PRBs} \leq 300$ and $n+n_{PRBs} \leq N$}
			\STATE $D_{PRBs}=D_{PRBs} + B_{n+n_{PRBs},u^*}$  
			\STATE $n_{PRBs}=n_{PRBs} + 1$
		\ENDWHILE
		\IF {$n+n_{PRBs} \leq N$}
			\STATE Update $\mathcal{U} = \mathcal{U}  \setminus  \{u^{*}\}$	
			\STATE $n=n+n_{PRBs}$	
			\STATE $C_{volte}=C_{volte}+253$
		\ELSE
			\STATE Find $k^*=\arg\max_{k\in\mathcal{K}} B_{n,U+k}$ 
			\STATE $D_{PRB}=B_{n,U+k^*}$ 
			\STATE $n=n + 1$
			\STATE $C_{data}=C_{data}+D_{PRB}$	
		\ENDIF											
	\ELSE  
		\STATE Find $k^*=\arg\max_{k\in\mathcal{K}} B_{n,U+k}$ 
		\STATE $D_{PRB}=B_{n,U+k^*}$ 
		\STATE $n=n + 1$
		\STATE $C_{data}=C_{data}+D_{PRB}$		
	\ENDIF		
\ENDWHILE  
\end{algorithmic}
\end{algorithm}{}

\begin{algorithm}{}
\caption{Heuristic Scheme for achieving PF} 
\label{alg:2}
\begin{algorithmic}[1]
\STATE Initialize: $n=1$, $C_k=0$ for all $k$
\WHILE {$n \leq N$} 
	\STATE Initialize: $D_{PRBs}=0$, $D_{PRB}=0$, $n_{PRBs}=0$	
	\IF {$\mathcal{U} \ne \emptyset$}
		\STATE Find $u^*=\arg\max_{u\in\mathcal{U}} B_{n,u}$ 
		\STATE $D_{PRBs}=D_{PRBs} + B_{n,u^*}$ 
		\STATE $n_{PRBs}=n_{PRBs} + 1$
		\WHILE {$D_{PRBs} \leq 300$ and $n+n_{PRBs} \leq N$}
			\STATE $D_{PRBs}=D_{PRBs} + B_{n+n_{PRBs},u^*}$  
			\STATE $n_{PRBs}=n_{PRBs} + 1$
		\ENDWHILE
		\IF {$n+n_{PRBs} \leq N$}
			\STATE Update $\mathcal{U} = \mathcal{U}  \setminus  \{u^{*}\}$	
			\STATE $n=n+n_{PRBs}$	
			\STATE $C_{volte}=C_{volte}+253$
		\ELSE
			\STATE Find $k^*=\arg\max_{k\in\mathcal{K}} (B_{n,U+k}/\overline{A}_k)$ 
			\STATE $D_{PRB}=B_{n,U+k^*}$ 
			\STATE $n=n + 1$
			\STATE $C_{k^*}=C_{k^*}+D_{PRB}$
			\STATE $C_{data}=C_{data}+D_{PRB}$	
		\ENDIF											
	\ELSE  
		\STATE Find $k^*=\arg\max_{k\in\mathcal{K}} (B_{n,U+k}/\overline{A}_k)$ 
		\STATE $D_{PRB}=B_{n,U+k^*}$ 
		\STATE $n=n + 1$
		\STATE $C_{k^*}=C_{k^*}+D_{PRB}$
		\STATE $C_{data}=C_{data}+D_{PRB}$		
	\ENDIF		
\ENDWHILE  
\end{algorithmic}
\end{algorithm}{}

\begin{figure}[htp]
\centering
\includegraphics[totalheight=53em]{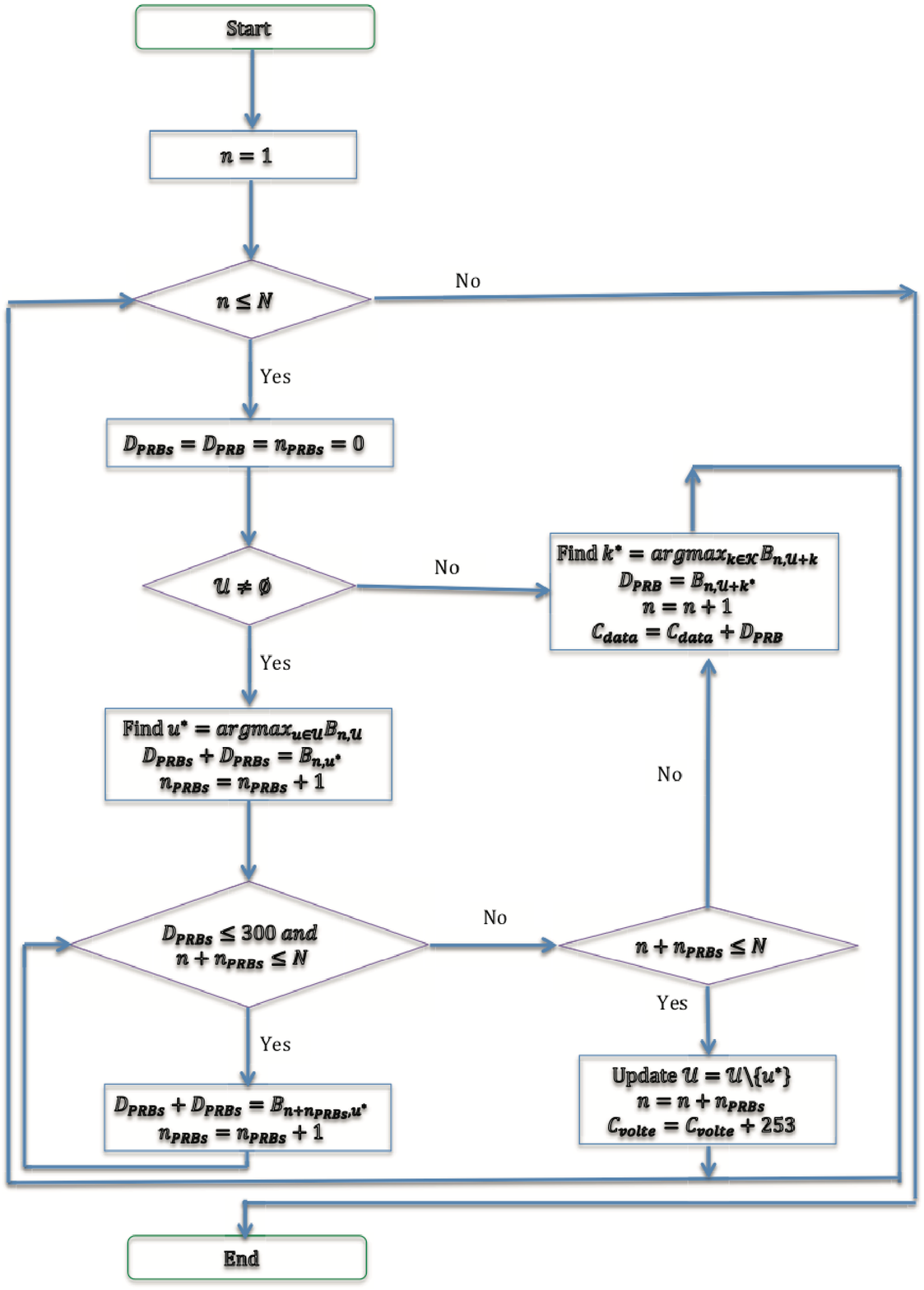}
\caption{Flowchart of Algorithm1}\label{fig:flowchart}
\end{figure}
The proposed scheme is detailed in Algorithm~\ref{alg:1}, and the flowchart of this algorithm is shown in Fig.~\ref{fig:flowchart}. We use $D_{PRBs}$ and $D_{PRB}$ to represent the number of bits that could be transmitted if the PRB were allocated to a VoLTE user or to a data user respectively. Define $n$ and $n_{PRBs}$, respectively, as the index of PRBs and the number of PRBs allocated to a VoLTE user. 

For each PRB, if $\mathcal{U}$ is non-empty, the VoLTE user which has the maximum SINR is selected as $u^*$. If the number of bits for that PRB ($D_{PRBs}$) is enough for one VoLTE packet (300 bits), that PRB ($n$) will be allocated to that user ($u^*$). Otherwise, we add the next PRB and we check if the number of bits for both PRBs are enough for one VoLTE packet. As soon as we reach the number of bits for a VoLTE packet, those PRBs will be allocated for the VoLTE user ($u^*$). If \emph{the number of PRBs are not enough for a VoLTE packet, the PRB will be assigned to the data user with highest SINR ($k^*$).} This is a crucial step wherein PRBs that are not useful to satisfy VoLTE users are allocated to data users. 

Each time after VoLTE user $u^*$ is scheduled to transmit, $C_{volte}$ is updated as
\begin{equation}
C_{volte} = C_{volte} + 253,
\end{equation}
and if the data user $k^*$ is scheduled to transmit, $C_{data}$ is updated as
\begin{equation}
C_{data} = C_{data} + D_{PRB}.
\end{equation}

As we will see in the next section, this remarkably simple scheme, an essentially greedy scheme, provides performance very close to that provided by the solutions to TTI-level problem and~\eqref{eq:obj1}-\eqref{eq:cons4}. For comparison, we will compare the results of Algorithm~\ref{alg:1} to those of a baseline scheduling algorithm in LTE in which the GBR users have strict priority over the non-GBR, data, users. So, the baseline scheme does not include steps 15-19 in Algorithm~\ref{alg:1}.

\subsection{Achieving PF Throughput}
As in the previous section, we now design a heuristic scheme to achieve proportional fairness amongst data users. The scheme is detailed in Algorithm~\ref{alg:2}, where the initial setting is the same as in Algorithm~\ref{alg:1}, except that $C_k$ is represents the throughput of data user $k$ in current TTI and is initialized to 0. The scheme is very similar to that in Algorithm~\ref{alg:1}. The main difference is that a weighted average rate ($\overline{A}_k$) is defined in Algorithm~\ref{alg:2} and initialized to 1 for all data users $k$. If all VoLTE users are scheduled to transmit or the remaining PRBs are not enough for one VoLTE packet, the data user with highest $B_{n,U+k}/\overline{A}_k$ will be selected to keep proportional fair throughput among data users. At the end of each TTI, $\overline{A}_k$ is updated as
\begin{equation}
\overline{A}_k = \gamma \overline{A}_k + (1-\gamma)C_k.
\end{equation}

\subsection{Complexity Analysis}

The heuristic schemes presented here are motivated by the complexity of the schemes in Sections~\ref{sec:optFr} and~\ref{sec:optTS}. With $N(U+K)T$ variables, the complexity of the frame-level optimization is $\mathcal{O}(2^{N(U+K)T})$. Similarly, the complexity of the schemes in Section~\ref{sec:optTS} are $\mathcal{O}(2^{N(U+K)}T)$.

In order to calculate complexity of the heuristics, we need to consider the worst case and evaluate the situation when values in the \texttt{if-else} conditions cause the maximum number of statements to be executed. In Algorithm~\ref{alg:1}, the worst case would be when all PRBs are allocated to data users as the number of PRBs are not enough for VoLTE packet. So, lines 1-11, 15-21, and 27-28 are executed.

When $N$, $U$, and $K$ are large, the complexity is due to the two \texttt{while} iterations (Lines 2 and 8) and also finding two $\arg\max$s (Lines 5 and 17). The complexity for performing $\arg\max$ is proportional to number of values being sorted. The complexity of algorithm is, therefore, $(\mathcal{O}(U)+\mathcal{O}(N)+\mathcal{O}(K))$ for each PRB, i.e., $\mathcal{O}(NU+NK+N^2)$ overall. The complexity analysis in Algorithm~\ref{alg:2} is similar to Algorithm~\ref{alg:1} and so its complexity is also $\mathcal{O}(NU+NK+N^2)$.

\section{Numerical Results}
\label{sec:results}
In this section, we simulate and compare the performance of different proposed scheduling schemes for both VoLTE and data users. The simulations are carried out for frequency-selective channels modelled by ITU for the extended typical urban channel~\cite{ETUCh}. In this model, channels are Rayleigh with 9 taps with relative power given in Table~\ref{table:ITUmodel}~\cite{ITUm}. Our simulations are performed for a frame of 20ms with an LTE bandwidth of 1.4MHz comprising 7 PRBs, 3 MHz (comprising 15 PRBs), and 10 MHz (comprising 50 PRBs). All users (both VoLTE and data) are uniformly distributed within a cell of radius 288 m. The path loss exponent is set to $\alpha = 3.8$. At each user, the SINR includes the interference caused by 18 cells of similar size surrounding the cell under test. Since all cells are assumed to be in the downlink, the SINR is independent of the PRB allocations in the neighboring cells. The simulation parameters are summarized in Table~\ref{table:sim}.
\begin{table}[h]
  \centering
   \caption{Power delay profile for Extended Typical Urban Model~\cite{ITUm}}\label{table:ITUmodel}
  \begin{tabular}{|c|c|c|c|c|c|c|c|c|c|}
    \hline
    {\bfseries Tap Number} & 1 & 2 & 3 & 4 & 5 & 6 & 7 & 8 & 9 \\
    \hline
    {\bfseries Tap Average Power (dB)} & -1.0 & -1.0 & -1.0 & 0.0 & 0.0 & 0.0 & -3.0 & -5.0 & 7.0 \\
    \hline
    {\bfseries Excess Delay (ns)} & 0 & 50 & 120 & 200 & 230 & 500 & 1600 & 2300 & 5000 \\
    \hline
  \end{tabular}
  \vspace{4pt}
 \end{table}

\begin{table}[t!]
  \centering
   \caption{Simulation parameters}\label{table:sim}
  \begin{tabular}{|c|c|}
    \hline
    {\bfseries Parameter} & {\bfseries Value} \\
    \hhline{--}
    Bandwidth & 1.4, 3, and 10 MHz \\
    \hhline{--}
    PRBs per TTI & 7, 15, and 50 \\
    \hhline{--}
    Simulation length (T) & 20 ms \\
    \hhline{--}
    Slot duration & 0.5 ms \\
    \hhline{--}
    Scheduling time (TTI) & 1 ms \\
    \hhline{--}
    Number of OFDM Symbols per Slot & 7 \\
     \hhline{--}
    Number of data users (K) & 5 and 50 \\
    \hhline{--}
    Cell radius & 288 m\\
    \hhline{--}
    Path loss exponent & 3.8 \\
    \hhline{--}
    Channel model & Extended Typical Urban Model \\
    \hhline{--}
  Number of simulation runs for each result & 30 \\
    \hhline{--}
  \end{tabular}
  \vspace{4pt}
 \end{table}

\begin{figure}
\centering
\includegraphics[totalheight=21em]{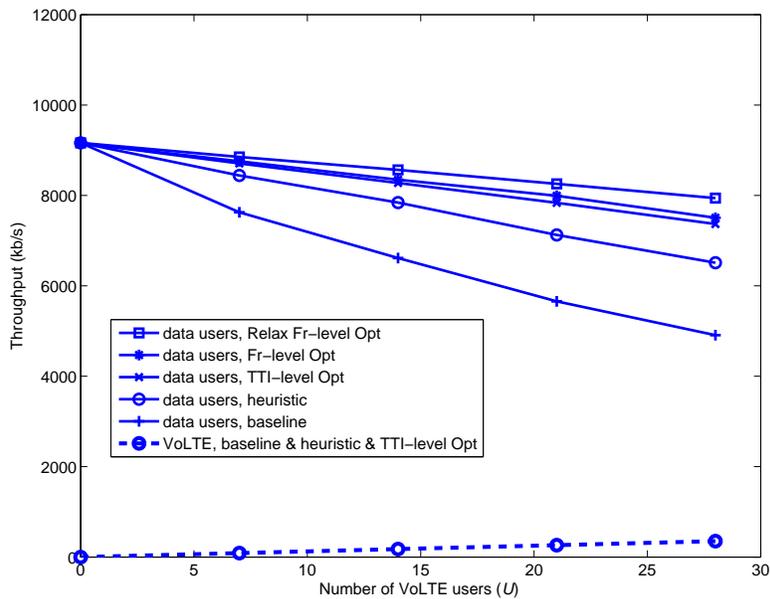}
\caption{Throughput for data users and VoLTE versus number of VoLTE users ($U$) for $K=5$}\label{fig:thoptM1}
\end{figure}

\begin{figure}
\centering
\includegraphics[totalheight=21em]{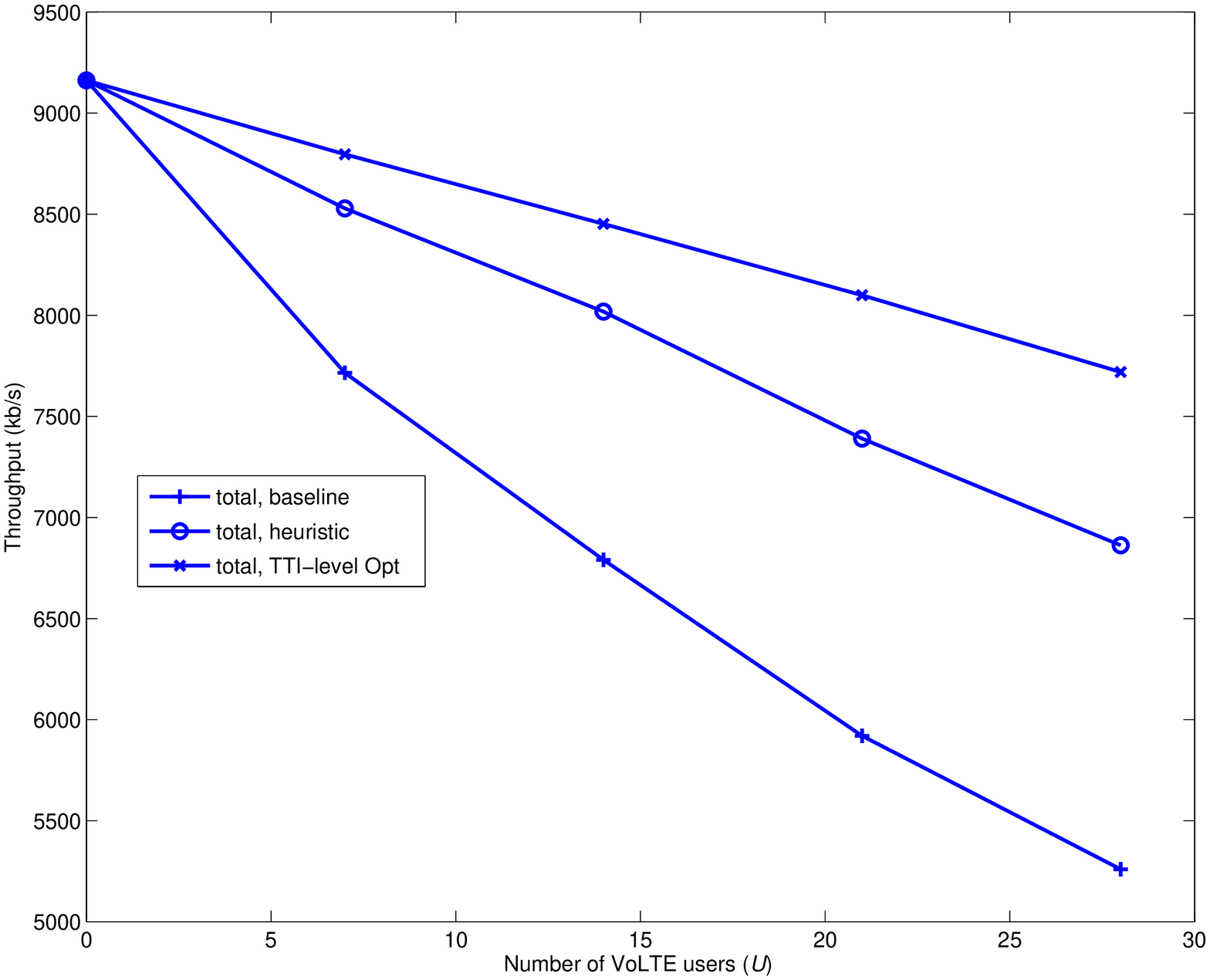}
\caption{Total throughput versus number of VoLTE users ($U$) for $K=5$}\label{fig:thoptM2}
		\centering
\includegraphics[totalheight=21em]{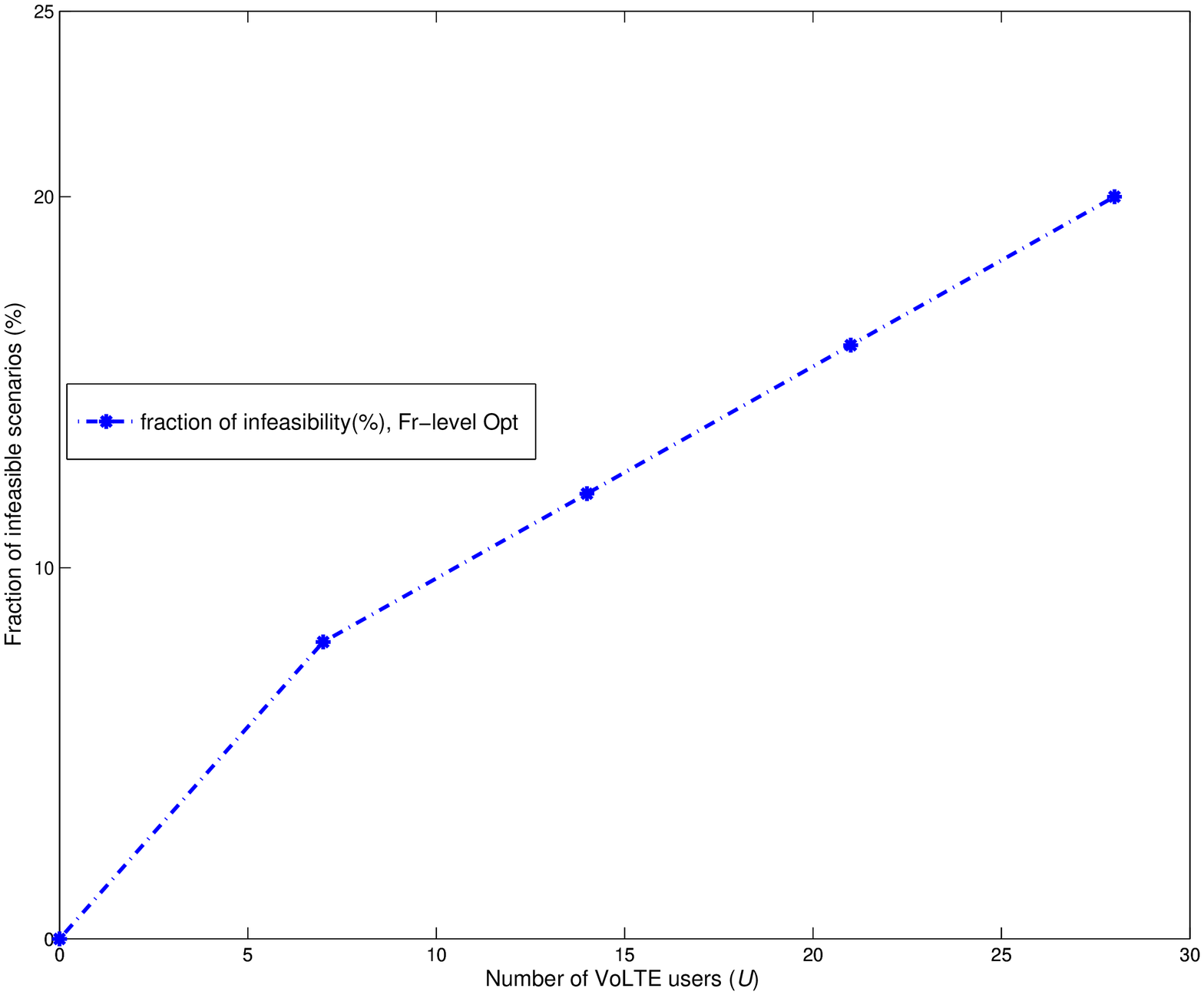}
\caption{Fraction of infeasible scenarios versus number of VoLTE users ($U$) for $K=5$}\label{fig:thoptM3}
\end{figure}

\begin{figure}[tp]
\centering
\includegraphics[totalheight=21em]{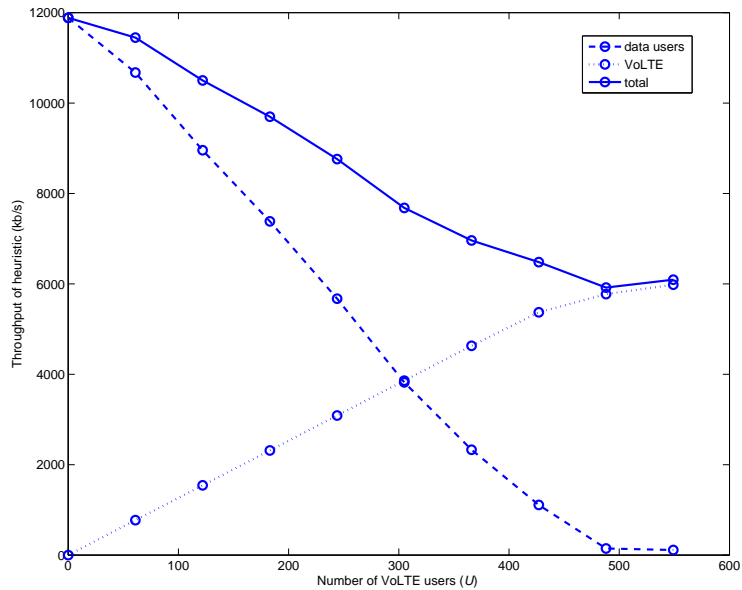}
\caption{Throughput versus number of VoLTE users ($U$)}\label{fig:thhe}
\end{figure}

\begin{figure}[htp]
\centering
\includegraphics[totalheight=21em]{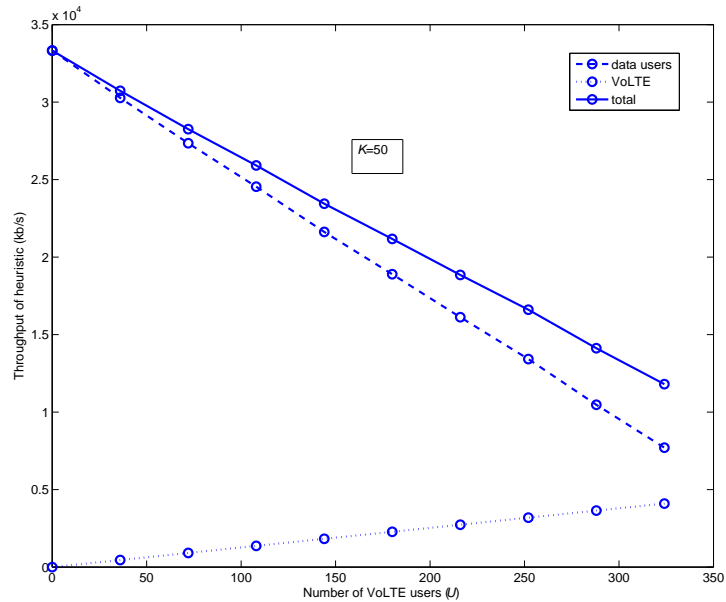}
\caption{Throughput versus $U$}\label{fig:thhe50d}
\end{figure}

Figs.~\ref{fig:thoptM1}-\ref{fig:thoptM2} and~\ref{fig:thhe}-\ref{fig:thhe50d} plot the throughput as a function of number of VoLTE users for various scenarios. We first present the results for 3 MHz bandwidth and $K=5$ data users in Figs.~\ref{fig:thoptM1}-\ref{fig:thoptM3}. Fig.~\ref{fig:thoptM1} plots the sum throughput for VoLTE and data users for the solution to the frame-level optimization problem, the TTI-level optimization problem and the heuristic scheme. As required, the VoLTE throughput is linear in the number of VoLTE users. Importantly, the data users' throughput performance of the proposed TTI-level optimization problem is very close to the frame-level optimum solution. 

The heuristic scheme provides much better throughput than the baseline scheme, which provides a strict priority to the GBR or VoLTE users. In other aspects, the baseline scheme is similar to the heuristic. The key gains are, therefore, due to the heuristic scheme intelligently assigning the PRBs to data users if the remaining PRBs are not enough for a VoLTE user. This figure also plots the results of the relaxed frame-level optimization (where the binary variables are relaxed to continuous variables). These results are close enough to the frame-level optimization results to be treated as a tight upper bound on performance. Also, there is a reasonable gap between the frame-level optimum solution and the heuristic scheme. There are two reasons for this gap: first, the frame-level optimization problem is over the whole frame (integer constraints are coupled across TTIs) while the heuristic scheme is executed in each TTI. Second, the frame-level optimum solution is based on the optimization framework while the heuristic scheme is the low-complexity algorithm.

This figure (and the others that follow) illustrates the importance of our scheduling algorithm. As the number of VoLTE users increases, a simplistic scheme results in an alarmingly large throughout loss for non-GBR data users. This is because the VoLTE traffic is, relatively, spectrally inefficient. In Fig.~\ref{fig:thoptM1}, a system with 28 VoLTE users results in a loss of approximately 43\% in the sum throughput. Developing a VoLTE-aware scheduler, such as proposed here, is essential to mitigate this large loss. This issue is underlined in Fig.~\ref{fig:thoptM2}, which plots the total throughput (sum over VoLTE and data users) for the  TTI-level optimization problem, the heuristic scheme and baseline scheme. As it is clear, the heuristic scheme (loss of 25\% as the number of VoLTE users increases from zero to 28) works much better than the baseline scheme (loss of 43\%) and its performance is relatively close to the TTI-level solution (loss of 16\%). The main reason for this gap is that the TTI-level solution is the outcome of two-phase integer optimization problems, which is a still high-complexity solution compared to the heuristic scheme. 

Finally, Fig.~\ref{fig:thoptM3} shows the percentage of times that the optimal solution is infeasible, i.e., at least one of the VoLTE users does not receive the required data rate. For the optimum method, we maximize the data users' throughput such that all VoLTE users are served. This figure illustrates one motivation for the TTI-level optimization; in cases of infeasibility, the frame-level problem does not lead to any solution - the problem is just declared infeasible.

The frame-level, and even the TTI-level, optimization problems are, essentially, impossible to solve beyond small values of $U$. However, the heuristic can be executed for a large number of VoLTE and data users. Figs.~\ref{fig:thhe}-\ref{fig:thhe50d} plot the sum rate for the VoLTE users, the non-GBR users, and total cell throughput for the heuristic scheme for an LTE bandwidth of 10 MHz. Fig.~\ref{fig:thhe} shows that when the number of VoLTE users increases, VoLTE throughput increases as VoLTE users are served with higher priority compared to non-GBR users. Consequently, the non-GBR throughput decreases since the radio resources (number of PRBs) is fixed. At the beginning, VoLTE throughput increases linearly which means that essentially all VoLTE users are served. Then, after about 430 VoLTE users, there are not enough PRBs to serve all VoLTE users so the sum rate of VoLTE users gradually saturates.

Fig.~\ref{fig:thhe50d} plots the throughputs achieved (VoLTE, non-GBR or data users and overall) when 50 data users ($K=50$) are available in a cell. As in Fig.~\ref{fig:thhe}, the heuristic is able to obtain a solution for a large number of VoLTE and data users. As before, the VoLTE users do place a significant burden on the system; the sum-throughput across all users falls significantly as the number of VoLTE users increases.

\begin{figure}[htp]
\centering
\includegraphics[totalheight=21em]{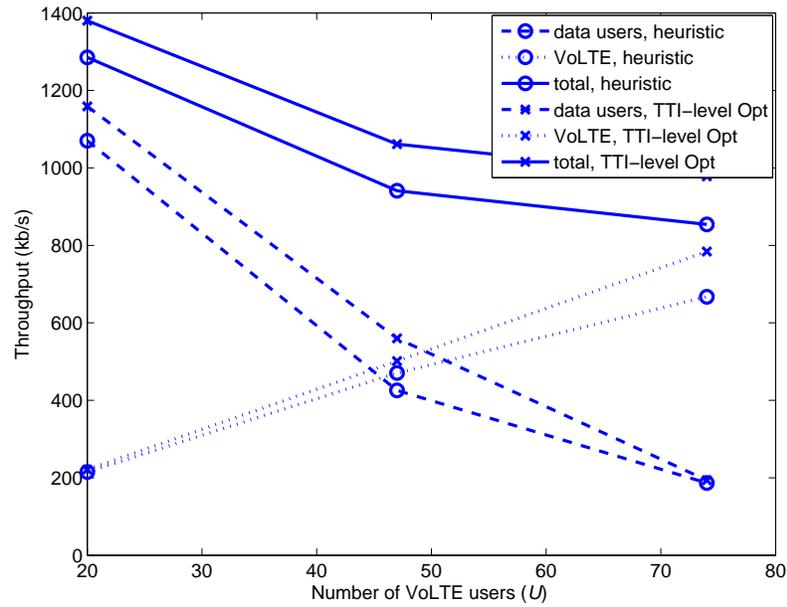}
\caption{Throughput versus $U$}\label{fig:thtshe}
\end{figure}

\begin{figure}[htp]
\centering
\includegraphics[totalheight=21em]{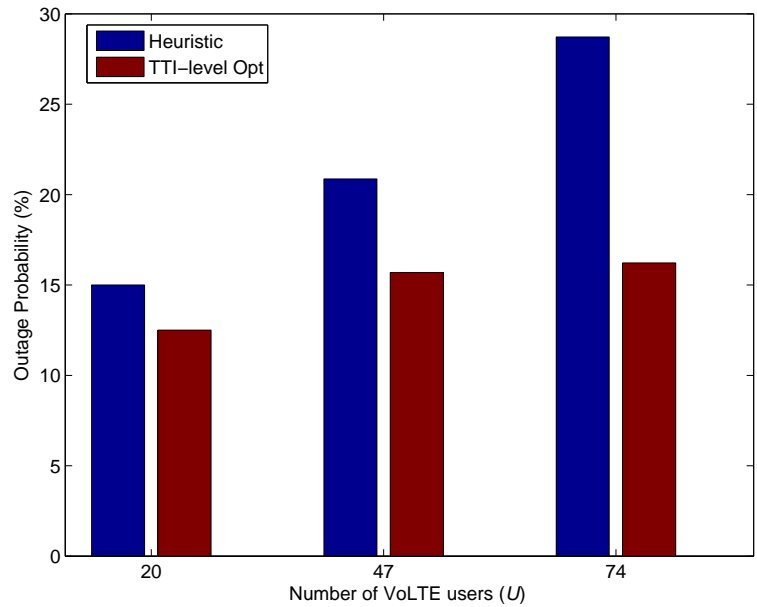}
\caption{Outage Probability versus $U$}\label{fig:outtshe}
\end{figure}

Figs.~\ref{fig:thtshe}-\ref{fig:outtshe} plot the throughput (VoLTE, non-GBR, total) and outage probability for VoLTE users when 1.4 MHz of bandwidth is available. At this low bandwidth, the penalty for using the heuristic is particularly clear. As we see, the TTI-level optimization outperforms the heuristic for both the VoLTE and data users. Fig.~\ref{fig:outtshe} confirms this trend in terms of the outage probability, i.e. the percentage of VoLTE users that are not served. As we expect, the outage probability for the heuristic method is larger and also for both methods it increases with number of VoLTE users. For small bandwidths, therefore, the TTI-level optimization algorithm may be worth pursuing.

\begin{figure}[htp]
\centering
\includegraphics[totalheight=21em]{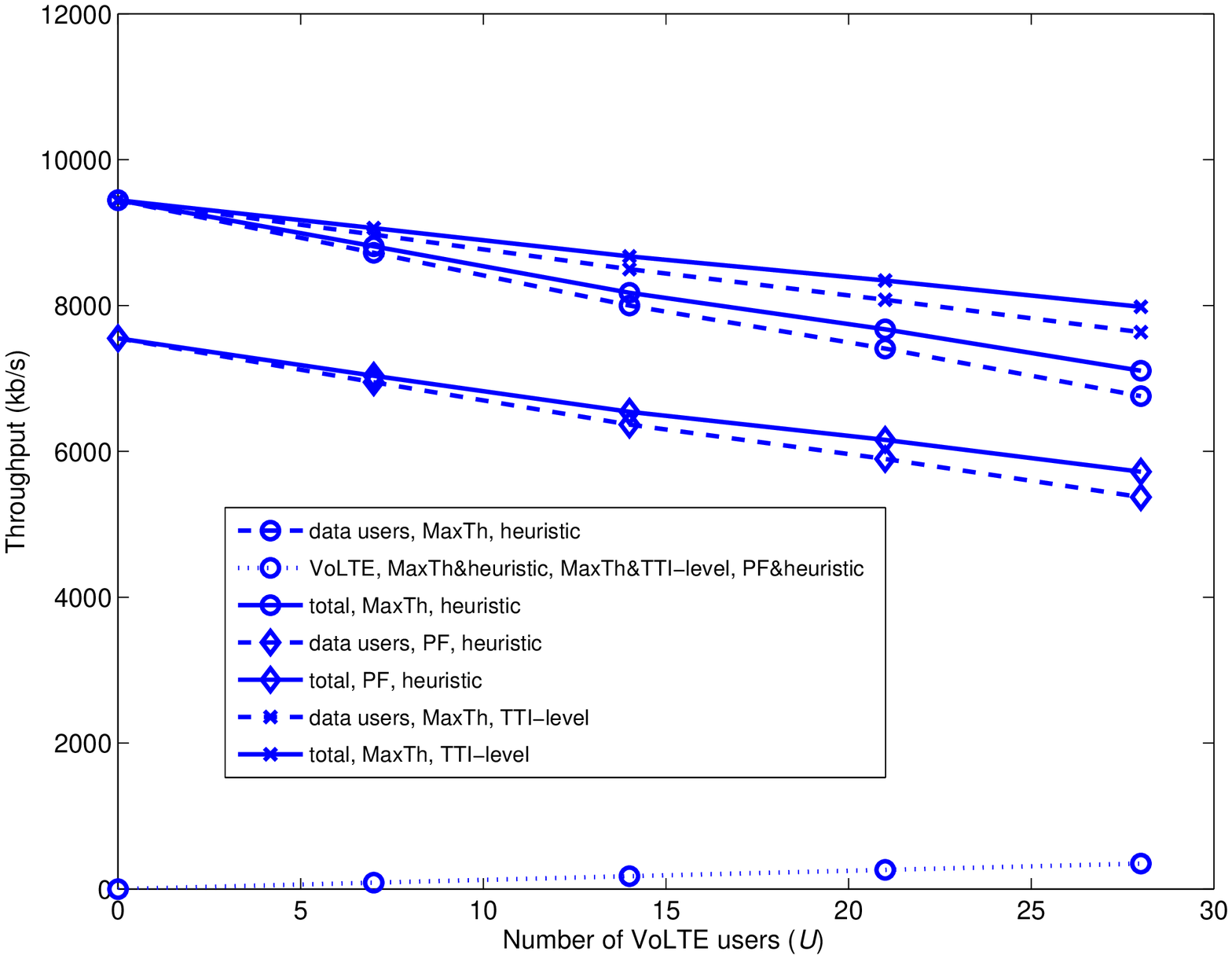}
\caption{Throughput versus $U$}\label{fig:thhepf}
\end{figure}

\begin{figure}[htp]
\centering
\includegraphics[totalheight=21em]{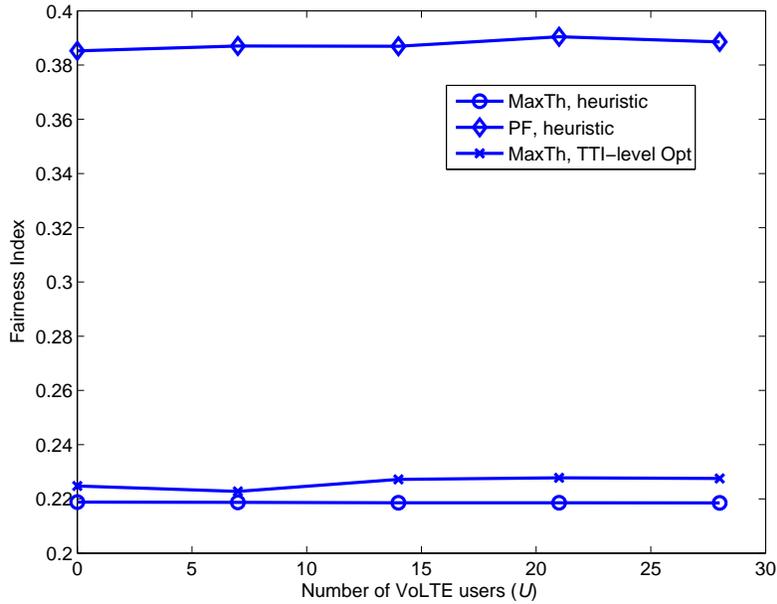}
\caption{Fairness Index versus $U$}\label{fig:fahepf}
\end{figure}

The final set of figures addresses the issue of fairness, implemented using the proportional fairness variants of the TTI-level optimization and heuristic. Figs.~\ref{fig:thhepf} and \ref{fig:fahepf} plot the throughput (VoLTE, non-GBR, total) and fairness index (Jain's fairness index) for the data users for an LTE bandwidth of 3 MHz. The plots show results for both objectives (maximizing total throughput and achieving PF). As expected, proportional fairness leads to a penalty in the sum throughput; importantly, the sum VoLTE throughput for both objectives is the same. Fig.~\ref{fig:fahepf} illustrates the gains by using PF: it plots the fairness index amongst the best-effort data users. As shown, the PF scheme achieves a far higher fairness index for the relatively low cost in  total throughput.

\section{Conclusions}
\label{sec:conc}
In this work, we developed scheduling and resource allocation to both VoLTE and data users in an LTE network. The objective is to maximize the total throughput while VoLTE users receive their required bit rate and with the associated timing. This approach improves on the baseline scheme wherein VoLTE users are scheduled first without accounting for their relatively poor spectral efficiency. To achieve our objective we developed two variants of the resource allocation problem: the first is a frame-level optimization which accounts for the details of the VoLTE application. However, this problem is extremely complex to solve. We therefore formulate a TTI-level optimization problem while yet meeting the VoLTE constraints. In addition, we propose a heuristic that significantly reduces computation time, at the cost in throughput. As our results show, the heuristic still far outperforms the baseline schemes.  

Our results show that, for the most part, VoLTE users can be satisfied. However, after a particular number of users, the VoLTE throughput saturates and all users cannot be satisfied. Importantly, while the frame-level optimization cannot handle infeasible cases, the TTI-level and heuristic provide service to most, but not all, users.

Our final formulation (and results) implements proportional fairness amongst the non-GBR data users. As expected, we achieve great improvements in Jain's fairness index for some cost in compare with maximum total throughput heuristic scheme.

A key contribution of this work is to develop optimization tools in the context of a wireless standard. Here, we chose to investigate VoLTE and data users; our approach illustrates new constraints and tools needed in an otherwise traditional sum-rate maximization problem.

\section*{Acknowledgement}
The authors would like to thank TELUS Canada and the National Science and Engineering Research Council of Canada for their financial support that made this research possible.
%

\end{document}